# Scaling Trajectories in Civil Aircraft (1913-1997)[•]

Koen Frenken[••]

*INRA/SERD, University Pierre Mendès-France*
*B.P. 47, 38040 Grenoble cédex 9, France*
*Tel.: 00.33.476.825.412; fax: 00.33.476.825.455; email: frenken@grenoble.inra.fr*

and

Loet Leydesdorff

*Science and Technology Dynamics, University of Amsterdam*
*Nieuwe Achtergracht 166, 1018 WV, Amsterdam, The Netherlands*
*Tel.: 00.31.20.525.6598; fax: 00.33.20.525.6579; email: l.leydesdorff@mail.uva.nl*

Using entropy statistics we analyse scaling patterns in terms of changes in the ratios among product characteristics of 143 designs in civil aircraft. Two allegedly dominant designs, the piston propeller DC3 and the turbofan Boeing 707, are shown to have triggered a scaling trajectory at the level of the respective firms. Along these trajectories different variables have been scaled at different moments in time: this points to the versatility of a dominant design which allows a firm to react to a variety of user needs. Scaling at the level of the industry took off only after subsequently reengineered models were introduced, like the piston propeller Douglas DC4 and the turbofan Boeing 767. The two scaling trajectories in civil aircraft corresponding to the piston propeller and the turbofan paradigm can be compared with a single, less pronounced scaling trajectory in helicopter technology for which we have data during the period 1940-1996. Management and policy implications can be specified in terms of the phases of codification at the firm and the industry level.

*Key words: aircraft, dominant design, scaling trajectory, probabilistic entropy, competence*

[•] The authors are grateful to Paolo Saviotti for the collaboration and his comments on earlier versions of this paper. The research reported here is partially based on data gathered in a project funded by the ESRC (Saviotti and Bowman 1984; Saviotti and Trickett 1993). The first author acknowledges financial support from the European Commission TMR-grant ERB4001GT961736; the second author is partially funded by TSER project PL97-1296.
[••] Corresponding author

1.      Introduction

Utterback and Abernathy (1975) have proposed the concept of a product life-cycle to describe technological evolution at the level of an industry.  At the start of a product life-cycle, a variety of product designs is being developed.  The competition between designs is eventually resolved into a dominant design.  Hereafter, innovation concentrates on process innovation and incremental product innovation given the dominant design.  Nelson and Winter (1977) and Dosi (1982) proposed to describe a series of incremental innovations within a stable design framework as a natural trajectory or technological trajectory, respectively.  Along a trajectory, development is guided and constrained by a set of heuristics which make up a technological paradigm.  The trajectory concept can be appreciated as the dynamic analogue of the concept of a dominant design.

Nelson and Winter (1977, 1982) and Sahal (1981, 1985) stressed that trajectories do not only concern periods during which the basic technological principles remain unchanged, but also a stage of incremental scaling of designs.  A prime example of a series of scaled models in civil aircraft has been the piston propeller DC-trajectory.  The scaling of the engine power, wing span, and fuselage length have led to improvements in speed by a factor of two, and in maximum take-off weight and range by a factor of five from the introduction of the DC3 in 1936 to that of the DC7 in 1956 (Miller and Sawers 1968; Jane's 1978).

The main heuristic of many technological paradigms can be represented in terms of the scaling of designs by means of step-by-step improvements.  However, specific design principles function only within a limited range of function levels, outside of which a structural redesign becomes necessary.  As Sahal (1985: 62) formulated it: *"(t)he point of departure of the theory advanced here is the well-known observation that change in size of an object beyond a certain point requires changes in its form and structure as well."*  Thus, technological development within a paradigm must come to an end at some point in time. This expectation justifies the idea that technological paradigms exhibit life-cycles.

For example, in the case of piston propeller aircraft technology, the life-cycle came to an end when it was understood that the functioning of propellers would decrease rapidly as cruising speed approached the speed of sound. Engineers envisaged that further scaling would become increasingly more difficult to realise due to non-linear rises in vibration and heat generation. Constant (1980) introduced in this context the concept of a *presumptive anomaly*. In the case of piston propeller technology the presumptive anomaly was circumvented by developing turbopropellers, rocket, jet, and turbofan engines as alternative technologies. The turbofan engine technology based on jet propulsion would eventually provide the basic technology for a new technological paradigm. Similarly, the quest for down-scaling computers led to a series of technological paradigms (vacuum tubes, transistors, integrated circuits).

A new paradigm has to compete during its initial phase with the momentum invested in the previous technology. A period of experimentation and recombination can be expected during which established firms may fall behind and new firms enter the market (Anderson and Tushman 1990). At this point, various firms follow their own trajectories which compete for dominance at the industry level. This competition process is resolved into a new dominant design which lays down a new set of heuristics. However, market segmentation may cause a bifurcation of trajectories as different design trajectories meet segregated market demands (Teubal 1979; Foray and Grübler 1990; Frenken *et al.* 1999). In the case of civil aircraft, turbofan aircraft with swept wings has become the dominant technology, but turbopropeller aircraft with straight wings is still used in short-range airline operations.

While single products can be expected to exhibit a life-cycle at the level of the firm, families of products are expected to exhibit a business cycle at the level of the industry (Foray and Garrouste 1991). The frequencies of these cycles are of different orders of magnitude (Simon 1969). The methodological problem is how to account for developments on both the micro-level and the macro-level. Below, we develop a scaling measure based on information theory which will enable us to analyse the data at the level of individual firms, at the industry level, and in relation to each other.

**2. Methods**

*2.1 Technical and service characteristics*

Statistical studies on patterns in technological change during the product life-cycle are based on a variety of data sources.[1] Following Saviotti and Metcalfe (1984), we shall use product characteristics to describe and compare product designs. The analysis is based on time-series of product characteristics for 143 civil aircraft used primarily for the transportation of passengers on a commercial basis (thus, cargo aircraft and business aircraft are excluded). The data was for a large part available from a previous project (Saviotti and Bowman 1984), and was extended to the year 1997. Data sources include the encyclopaedia of Jane's (1978, 1989, 1998), Chante (1990), and Green and Swanborough (1982). A description of the characteristics is provided in *Table 1*.

---

[1] These data sources include prices, sales, entry-exit and patent data (Gort and Klepper 1982), innovation counts, sales, entry-exit and productivity data (Klepper and Simons 1997), product characteristics, sales and entry-exit data (Tushman and Anderson 1986; Anderson and Tushman 1990), patent data (Malerba and Orsenigo 1996), and product characteristics (Sahal 1981, 1985; Saviotti 1996; Frenken *et al.* 1999).

**Table 1**
*Description of data on civil aircraft*

```
Number of cases:  143
Time period:  1913 – 1997
Scope:  all countries

Product characteristics          Unit of measurement

(technical)
1. engine power                  Kilowatt
2. wing span                     Meter
3. fuselage length               Meter

(service)
4. take-off weight               Kilogram
5. speed                         Kilometer per hour
6. range                         Kilometer
```

Saviotti and Metcalfe (1984) distinguished between technical and service characteristics of products. Technical characteristics were defined as variables that can directly be manipulated by producers (e.g., engine power). Variables that users take into account in their purchasing decisions (e.g., speed), were considered as service characteristics. Producers attempt to raise the product's service characteristics by manipulating technical characteristics, while users express their wants through the formulation of a set of service characteristics and their values.[2] Product designs then, can be considered as "interfaces" between supply and demand (Simon 1969). These interfaces can be represented in terms of trade-offs between technical and service characteristics. Technological innovations then, can be considered as improvements in these trade-offs.

The process of product design contains an ongoing search between users and producers for an optimal

---

[2] The characteristic approach to technological innovation can be considered an adaptation of Lancaster's (1966) demand approach. While Lancaster needed only a set of service characteristics, a requires a

match between technical and service specifications. In the case of product innovation, uncertainty prevails on the side of both the producer and the user (Clark 1985; Andersen 1991). Users are expected to select particular designs on the basis of functions, but new user wants can be envisaged when new combinations of technical characteristics become possible. Therefore, the assumption of a stable market environment selecting upon a variation of techniques solely on the basis of prices, can no longer be taken for granted. For example, the engine power needed to fly an aircraft of a certain size is bounded by technical principles, but its fuel consumption bears on the operating costs taken into account by users. The relative technical and commercial importance of engine power is thus at variance over time. The distinction between technical and service characteristics is not always easy to apply. The scaling measure as developed below takes into account all relations between characteristics, whether defined as a technical or a service characteristic, thus solving the definition problem without losing its conceptual meaning.

Scaling trajectories can be expected when the definitions of and the relationships between various characteristics of a product become stabilised ("closure"). Innovations aiming at scaling may be motivated by signals internal to the technology; the scaling in some parts of a technology can generate imbalances in other parts which in turn call for adjustments (Rosenberg 1969). During scaling processes, however, the overall design architecture and the set of functions of a technology remain largely invariant over prolonged periods of time. Rather, technical and service characteristics can then be expected to co-evolve (Windrum and Birchenhall 1998). A quantitative empirical analysis should enable us to indicate these periods of relative stability.

2.2     *A dynamic distance measure based on information theory*

---

representation of the supply side as a set of technical characteristics. See also, Saviotti (1996).

Designs make up the interface between supply and demand as it is expressed in the various trade-offs among characteristics. Scaling then, is indicated when these trade-offs remain stable over time. We will model the trade-offs by using the ratios among all six characteristics, thus taking into account both the relations among and between technical and service characteristics. Scaling may affect all these ratios. If none of the trade-offs is changed, the scaling is not innovative.

For example, for the product characteristics of the Douglas DC3 we have: *engine power* = 1636 kwatt, *wingspan* = 28.96, *fuselage length* = 19.63, *take-off weight* = 12701 kg, *speed* = 274 km/h, and *range* = 1650 km. The thirty ratios starting from engine power/wingspan till range/speed are:

| | | | | |
|---|---|---|---|---|
| 56.4917127 | 83.3418237 | 0.1288088 | 5.9708029 | 0.9915152 |
| 0.0177017 | 1.4752929 | 0.0022801 | 0.1056934 | 0.0175515 |
| 0.0119988 | 0.6778315 | 0.0015455 | 0.0716423 | 0.0118970 |
| 7.7634474 | 438.5704420 | 647.0198675 | 46.3540146 | 7.6975758 |
| 0.1674817 | 9.4613260 | 13.9582272 | 0.0215731 | 0.1660606 |
| 1.0085575 | 56.9751381 | 84.0550178 | 0.1299110 | 6.0218978 |

The set of ratios can be considered as a probability distribution *(p₁,...,p₃₀)* by dividing each ratio by the sum of the ratios. In this manner, we obtain a probabilistic representation for each aircraft.

In order to analyse the scaling development in subsequent product designs, one is in need of a distance measure between the representations in terms of probability distributions. Using information theory, one is able to calculate on the basis of a distribution the *expected information content* contained in the message that the distribution has changed as a next design was introduced on the market using the following formula (Theil 1967, 1972):

$$I(q \mid p) = \sum_{i=1}^{30} q_i \log_2 (q_i / p_i) \tag{1}$$

The *expected information content* of the *a posteriori* distribution ($q_1,...,q_{30}$) given the *a priori* distribution ($p_1,...,p_{30}$) can also be considered as an information-theoretical distance between product designs in terms of scaling. If none of the trade-offs was changed, the probability distribution has remained the same. Compared with the previous product, such a design would be a perfectly scaled version. In that case, every $q_i$ is equal to its corresponding $p_i$, so that *I* vanishes, since $log_2(1) = 0$. It can be shown that otherwise *I* is positive (Theil 1972, pp. 59f.): the message that change has occurred is expected to contain information or, in other words, a probabilistic entropy is generated.

In the following, we shall use *I* as a measure of the degree of scaling between two product designs: *the lower the value of I, the more similar are the ratios between two product designs and the more the latter design can be considered as a scaled version of the former design*. Note that our approach differs from that of Sahal's (1981, 1985) who used parametric tests on scaling constants over long periods of time. We do not assume that scaling has a unequivocal direction throughout the process of scaling.

For example, if we want to compare the distribution of ratio values of the Douglas DC4 with the Douglas DC3, we take the DC3 as the *a priori* distribution ($p_1,...,p_{30}$), and the DC4 as the *a posteriori* distribution ($q_1,...,q_{30}$). As done for the DC3 above, one is able to calculate the ratios between the product characteristics of the DC4, and divide these ratios by the sum of ratios to obtain the envisaged representation of the DC4. Using formula (1), we can then calculate the scaling distance between these two product designs, that is, the *I (DC4 | DC3)*.

*2.3    A measure of critical transition*

The advantage of the algorithmic approach becomes clear when we compare three instances in a series like the sequence of three products A-B-C. In a geometrical representation of differences between three products designs in an Euclidean space, the distance between A and C will be smaller than the sum of the distances between A and B and between B and C (Theorem of Pythagoras). In contrast, the information-theoretical distance between two products A and C being $I(C|A)$ is not necessarily smaller than the sum of the distance between products A and B being $I(B|A)$ and between products B and C being $I(C|B)$.

**FIGURE 1 ABOUT HERE**

Three designs A, B, and C, and their respective distances *I* are depicted in *Figure 1*. Design A precedes B, and B precedes C. Using the dynamic information measure, it is possible that the sum of the intermediate distances between A and B, and B and C, will be smaller than the distance between A and C. In this case:

$$I(B|A) + I(C|B) < I(C|A) \qquad (2)$$

This formulation is equivalent to:

$$I(B|A) + I(C|B) - I(C|A) < 0 \qquad (3)$$

If this inequality is confirmed, the transition from design A to design C via design B can be considered a *critical transition*.

The inequality enables us to evaluate the function of the intermediary. In the normal case, one expects an intermediate design B in a series A-B-C to improve the prediction of design C in comparison to a previous design A, that is, $I(C|B) < I(C|A)$. In the case of a *critical* transition, however, the *sum* of the intermediate informational distances ($I(C|B) + I(B|A)$) is smaller than the informational distance between design A and design C ($I(C|A)$). From the perspective of design C, the "signal" sent by design A was amplified by design B as an auxiliary transmitter to such an extent that design B "boosts" the signal from A (Leydesdorff 1992, 1995). Thus, the transition is critical for the further path of the development.

We use the term "critical transition" rather than "path-dependent transition", because the concept of path-dependency was defined by Arthur (1989) in relation to the non-ergodic technological development in networks of adopters. We measure historical series of individual designs (David 1985; cf. Foray 1998), where a critical transition holds that a single design has reoriented the scaling pattern into a *new* direction. Below, we apply this measure to sequences in aircraft designs, taking the Douglas DC3 and the Boeing 707 as intermediate designs, since these two designs have been identified as dominant designs (e.g., Constant 1980; Gardiner 1986b).

*2.4    Scaling measurement at the level of the industry*

The probability distribution of ratios provides us with a unified representation of each product design in the dataset. This enables us to measure scaling patterns in a series of product designs. For example, we will analyse below the scaling trajectories in the Douglas and Boeing companies. The trajectory notion, however, also refers to an industry-wide convergence in scaling patterns. In order to measure scaling trajectories at this level of aggregation, one needs to compare each product design with all other designs relevant in the competition.

Although each observation maintains a distance from all other observations in the dataset, not all observations are relevant for the comparison. We are interested in the development and competition in certain periods. The population thus has to be determined dynamically. The data show that the major development cycles of individual firms have been the DC3-MD11 series with an average frequency of a new product each 5.6 years, the Boeing 707-777 series with an average of 6.2 years, and the Fokker F-series with an average of 8.0 years. Since pre-war product development cycles were usually considerably smaller than post-war cycles, we have simplified our computations by using a time window of five years after the year of introduction. Thus, we compare the design of an aircraft introduced in, say, 1936, with all designs introduced during the period 1937-1941. However, we have tested our results for their robustness by changing the time-frame to ten years (e.g., 1937-1946). This did not affect our conclusions.

On the one hand, a test of scaling at the industry level needs to take into account the rate of *diffusion* of the ratios between product characteristics. On the other hand, the codification of design principles associated with the emergence of a dominant design also implies a *convergence* of particular design principles that have been developed in the past (Dosi 1982). Thus, the coming into existence of a scaling trajectory at the industry level is essentially a two-sided phenomenon. It refers both to the diffusion of design principles, and to the convergence of design principles. These phenomena are different: *the diffusion of particular design principles does not necessarily imply convergence of design principles, since a design can be scaled in various different and potentially divergent directions.* For example, some aircraft firms may scale a dominant design with respect to maximum take-off weight, others with respect to speed, and still others with respect to range. Hence, to test the dominant design hypothesis, one needs to distinguish between the diffusion of design principles through time and the convergence of design principles that can be observed in retrospect.

In our information-theoretical framework, these different dynamics call for a change in the *a priori* and the *a posteriori* frames of reference. In the case of diffusion of a design through time, the frame of reference is a particular product design as the *a priori* expectation of future designs. The diffusion of a product design can then be measured by its distance *I* following formula (1) to all the members of the technological population as *a posteriori* events at next moments in time (*Figure 2a*). The average of *I*-values is then obtained by dividing the sum of *I*-values by the number of comparisons *N* during the respective five years of observation. This average value indicates the extent to which a design has diffused throughout the industry. Diffusion here refers to the subsequent scaling of a particular product design, and not the diffusion of an aircraft in terms of its sales. Remember that $I(q \mid p)$ is an inverse indicator: *a low I-value indicates a high degree of diffusion of a product design, while a high I-value indicates a low degree of diffusion*.

The hypothesis that a scaling trajectory is characterised by common heuristics among firms, holds that ratios between product characteristics will tend to converge. The degree of convergence, then, is indicated by the distances between all the products in a population that preceded a product, when taking the latter as the frame of reference with hindsight (*Figure 2b*). Thus, the single design introduced later in time is taken as a yardstick, while the population of products preceding this design during five years is considered the set of events relevant for the evaluation. The average of the *I*-values indicate the degree of convergence, where *a low I-value indicates a high degree of convergence and a high I-value indicates a low degree of convergence*.

**FIGURE 2 ABOUT HERE**

### 3. A stylised history of civil aircraft

Before turning to the computational results, let us briefly return to the expectation. We have used the civil aircraft industry since the history of its technological developments is one of the best documented. Both historians and economists have analysed these technological development in considerable detail (Miller and Sawers 1968; Constant 1980; Mowery and Rosenberg 1982; Sahal 1985; Gardiner 1986a, 1986b; Vincenti 1990; Bilstein 1996). Although these studies differ in perspectives and in the levels of their quantitative and qualitative measurements, there is general consensus on the following findings:

1. The early history of civil aircraft, covering roughly the period between 1920 and 1935, is characterised by a large variety of designs and a small market demand. The main bottleneck concerned the limited range of flight, which rendered the competitiveness of this technology low compared to the services of trains and boats.

2. In the thirties, new product designs in the United States allowed for long-range airline operations at reasonable speed levels. As a result, aircraft out-competed trains in terms of speed, comfort, and price over long distances (e.g., coast-to-coast flights). The Douglas DC3, an all-metal, monocoque piston propeller monoplane with engines placed under the wings, is commonly considered as the dominant design. Because of the rapid increase in market demand for cheap air-traffic, learning curves of the DC were particularly steep. The total production of the DC3 was over 10,000 models, including military production (Jane's 1978).

3. During the fifties, market demand in European countries increased rapidly. Some European producers followed the successful American piston propeller aircraft, while other firms used turbopropeller aircraft (e.g., Vickers, Fokker). Early attempts of DeHavilland and Aerospatiale to commercialise jet engine technology in civil aircraft failed.

4. In the late 1950s, a second revolution in civil aircraft design took place as the Boeing company

successfully introduced the Boeing 707, a long-range turbofan aircraft that had been redesigned from this company's bomber line. The introduction of the Boeing 707 was followed by a series of turbofan aircraft labelled the 700-series. In the late sixties, McDonnell Douglas and Lockheed succeeded in introducing long-range turbofan aircraft at competing price levels. Only a few European and Sovjet aircraft designers developed civil turbofans, while the majority focused on short-range turbopropeller planes.

5. Finally, during the seventies and eighties, turbofan technology diffused throughout the market for medium and long-range distance flights. Most recently, within the segment of short-range aircraft which has traditionally been covered by turbo-propeller aircraft, turbofan technology has also penetrated, rendering the diffusion of this new paradigm nearly complete (Jane's 1995).

In summary, two major breakthroughs have taken place which revolutionised the civil aircraft industry. The advent of the Douglas DC3 in the mid-thirties, which opened up a mass market and imposed a dominant design in piston propeller aircraft, and the introduction of the Boeing 707 in the late fifties, which radically extended payload, speed and range capabilities using turbofan engine technology and swept wings. The production and operating costs of these designs and their subsequent follow-up models (the DC-series and the 700-series, respectively) decreased rapidly as a result of high sales figures which allowed for steep learning curves. Both the DC3 and the Boeing 707 are said to have reoriented design principles at the industry-wide level. In the remainder of this article we will focus on the questions of whether these two designs have generated critical transitions, and whether their introduction forced the industry to follow a common scaling trajectory.

**4.     Critical transitions at the firm level**

*4.1     The Douglas DC3*

Using the inequality given in formula (3), product sequences including the Douglas DC3 as the intermediate design (*B* in Figure 1) and the Douglas DC4 as the follow-up design (*C*) were used for testing whether the DC3 marked a dominant design for the Douglas company. As preceding designs (*A*), we selected thirteen historically important precursors including the Boeing 247D and Fokker's Trimotor, which have been identified as the DC3's major competitors at the time (Miller and Sawers 1968). *Table 2* shows the results of the tests on critical transitions in product sequences that include the DC3 and the follow-up DC4 model, as defined in formula (3) above. The negative values indicate critical transitions, while positive values indicate non-critical transitions. The results show that the DC3 constituted a critical transition with respect to all aircraft designs preceding the DC3 *within the United States*. For some European aircraft models, the test do not indicate a critical transition. The results show, however, that the DC3 has reoriented the design principles for the Douglas company as compared to previous aircraft models within the U.S. market.

**Table 2**
*Test for critical transitions in the case of the Douglas DC3*

| product sequence | country | value (in bits) |
| --- | --- | --- |
| Lockheed Vega (5B) - DC3 - DC4 | US | -0.0354 |
| Douglas M.4 – DC3 - DC4 | US | -0.0267 |
| Northrop Delta - DC3 - DC4 | US | -0.0247 |
| Lockheed Electra - DC3 - DC4 | US | -0.0191 |
| Ford Trimotor - DC3 - DC4 | US | -0.0165 |
| Boeing 247D - DC3 - DC4 | US | -0.0138 |
| Boeing M.80 - DC3 - DC4 | US | -0.0033 |
| Fokker F.10 Trimotor - DC3 - DC4 | US/Netherlands | -0.0128 |
| Armstrong Whitworth Argosy - DC3 - DC4 | UK | +0.0034 |
| Junkers J.52 - DC3 - DC4 | Germany | -0.0035 |
| Heinkel HE 111 - DC3 - DC4 | Germany | +0.0018 |
| Potez 62 (O) - DC3 - DC4 | France | -0.0022 |

| Bloch M.B. 220 - DC3 - DC4 | France | +0.0057 |

We also tested the chronological DC/MD-series which succeeded the emergence of the DC3 for critical transitions. These results are listed in *Table 3*. Except for DC7-DC8-DC9 sequence, no critical transitions have taken place. The DC8 implied a fundamental reorientation of design principles, since from then (1967) onwards, Douglas used turbofan engines instead of piston propeller engines. Obviously, turbofan technology implied a set of new design principles (including swept wings which allowed for a smaller wing span in relation to fuselage length). The ratios between product characteristics hence were organised in a new manner and scaled according to new heuristics. In the history of the Douglas company, we can thus distinguish a piston propeller and a turbofan trajectory.

During all other transitions, no reorientation in design principles took place within the series. All models developed before the DC8 concerned piston propeller aircraft, which all built upon the DC3-design, but in different respects. For example, the DC4 was a up-scaled version of the DC3, while the DC5 was a down-scaled version of the DC3. The DC6 was a faster version of the DC4 made possible by its pressurised cabin which allowed for higher altitudes. The DC7 was marked by its long range. The models developed after the DC8 concern turbofan aircraft. Again, scaling pursued different directions from the DC8 onwards: the DC9, MD 80 and MD 85 are all down-scaled aircraft in all product dimensions. The DC10 is also smaller than the DC8 but has a higher engine power and maximum take-off weight. Finally, the MD11 is the largest aircraft developed by McDonnell Douglas exceeding all previous aircraft in terms of engine power, length, maximum take-off weight and speed.

**Table 3**
*Test for critical transitions along the Douglas DC-trajectory*

| DC3 – DC4 – DC5 | +0.0290 |
| DC4 – DC5 – DC6 | +0.1710 |
| DC5 – DC6 – DC7 | +0.0136 |
| DC6 – DC7 – DC8 | +0.0037 |
| DC7 – DC8 – DC9 | -0.0176 |
| DC8 – DC9 – DC10 | +0.0251 |
| DC9 – DC10 – MD80 | +0.0463 |
| DC10 – MD80 – MD11 | +0.0407 |
| MD80 – MD11 – MD85 | +0.0284 |

The results show that along both the piston propeller trajectory DC3-DC7 and the turbofan trajectory DC8-MD85, various characteristics have been scaled at different stages and in different directions. The (McDonnell) Douglas company could thus react to shifting technical problems and market opportunities while building on a set of design principles which only once has undergone a critical change. This dynamic can be related to Rosenberg's (1969) notion of shifting technological imbalances along a technological trajectory which direct research efforts as focusing devices. Furthermore, the versatility of a dominant design allows a firm to meet a variety of user needs by means of scaling the design into different directions. The robust nature of a dominant design is an important source of competitive advantage over other firms (Clark 1985).[3]

*4.2   The Boeing 707*

In 1957, Boeing introduced the 707 as a new civil aircraft using turbofan engine technology. This aircraft has allegedly served as a dominant design for the subsequent 700-series. Using our methodology, we tested product sequences including the Boeing 707 as the dominant design (*B* in Figure 1) and the Boeing 727 as the follow-up design (*C*).[4] As preceding designs (*A*), we used the latest developments of that time in piston propeller aircraft (Boeing 377, Douglas DC7, Lockheed Constellation), state-of-the-art models in turboprops (Vickers Viscount, Bristol Britannia), and early jet and turbofan aircraft (Aerospatiale Caravelle, DeHavilland Comet). The results are listed in *Table 4*. All precursors are found to yield critical transitions when compared with the Boeing 707 and its successor the Boeing 727. Again, the results confirm the hypothesis that the allegedly dominant design marked a critical transition for the Boeing company. Contrary to the results for the DC3, we find that the Boeing 707 implied a reorientation with respect to *all* precursors considered here, irrespective of the country of origin.

---

[3] Confer Gardiner's (1986b : 143) distinction between *robust* and *lean* designs : "a robust design is one that brings together several new divergent lines of development to form a new 'composite' design, which is then internally adjusted to form a new 'consolidated' design, which is then further developed as a variety of 'stretched' design. Lean designs fail at one, two, or more often all three of these stages". Gardiner mentions the DC3 and the Boeing 707 as examples of robust designs in the history of civil aircraft.

[4] The Boeing 717 has never been used for civil services.

**Table 4**
*Test for critical transitions in the case of the Boeing 707*

| product sequence | Country | value (in bits) |
|---|---|---|
| Lockheed Constellation - 707 – 727 | US | -0.0305 |
| Boeing 377 Stratocruiser - 707 – 727 | US | -0.0297 |
| Douglas DC7 - 707 – 727 | US | -0.0180 |
| Bristol Britannia - 707 – 727 | UK | -0.0187 |
| De Havilland Comet - 707 – 727 | UK | -0.0157 |
| Vickers Viscount - 707 – 727 | UK | -0.0093 |
| Aerospatiale Caravelle - 707 – 727 | France | -0.0083 |

Testing the trajectory development *within* the Boeing 700-series (*Table 5*), we found that only the most recent transition is critical, while all other transitions were found non-critical. The result for the most recent transition suggests that it was not the development of the 707, but the development of the 767 that laid the basis for the 777. Indeed, the 777, the largest two-engine aircraft ever built, elaborates on the two-engine lay-out incorporated in the 757 and 767, as different from the four-engined Boeing 747. As was the case in Douglas DC3, Boeing's dominant 707 design has been elaborated in various directions. The 727 and the 737 were down-scaled models using three and two engines, respectively, instead of the four-engine set-up of the 707. The 747 is a up-scaled wide-body version with four engines, while the 757, 767 and 777 all incorporated two engines.

**Table 5**
*Test for critical transitions along the Boeing 700-trajectory*

| | |
|---|---|
| 707 – 727 – 737 | +0.0362 |
| 727 – 737 – 747 | +0.0676 |
| 737 – 747 – 757 | +0.0349 |
| 747 – 757 – 767 | +0.0155 |
| 757 – 767 – 777 | -0.0026 |

## 5. Diffusion and convergence at the industry level

In the previous section, we have analysed developments in aircraft design in terms of product sequences. The results indicate that both the DC3 and the Boeing 707 reoriented design principles when compared with their precursors. The results so far suggest only that the DC3 and the Boeing 707 have functioned as dominant designs at the *firm* level. These designs, however, are not necessarily the dominant designs that diffused throughout the industry. This question must be analysed in terms of dynamics at the level of the entire population of products.

### 5.1 The results on diffusion

As explained in *section 2*, one is able to calculate an average value of the expected information content for each product design in relation to relevant successive designs ("diffusion") and in relation to its precursors ("convergence"). As an indicator of the rate of the diffusion of a design, the average *I*-value was computed for each product with respect to all products introduced during the five years after its introduction. Note again that a product with a low *I*-value refers to a product design which has diffused to a large extent in the sense that its trade-offs have been scaled in succeeding products. A product with a high *I*-value has not diffused in the sense that succeeding designs have used rather different trade-offs. Note also that diffusion here concerns the subsequent scaling of a particular product design, and is not to be confused with the diffusion of an aircraft in terms of its sales.

**FIGURE 3 ABOUT HERE**

The results on the diffusion values are plotted in *Figure 3*. Each point refers to one aircraft model. The curvature suggests two cycles, each consisting of two stages: an experimentation stage with high *I*-values (that is, a low rate of diffusion) and a diffusion stage with low *I*-values (a high rate of diffusion). These results correspond to the cyclic dynamic of scaling trajectories, which are associated with particular technological paradigms as discussed above (cf. Constant 1980).

During the first cycle of piston propeller aircraft technology, the 1942-Douglas DC4 (*I=0.024*) and the 1944-Boeing 377 'Stratocruiser' (*I=0.028*) are among the smallest values indicating their importance at the industry population level. Both these designs concern four-engined piston propeller aircraft that followed upon the two-engine versions of the Douglas DC3 and the competing Boeing 247, respectively. Thus, although the Douglas DC3 is widely considered to have laid down the dominant design principles, it is the redesigned, four-engine version that diffused globally throughout the world market.

With respect to the second cycle of aircraft with turbofan engines, the Airbus-line A300-A310-A320 introduced during the eighties is among the lowest values (*I=0.046*, *I=0.046*, *I=0.034*, respectively). These Airbus models were introduced during the eighties to compete with the Boeing 757 introduced in 1982 and the Boeing 767 introduced in 1981, which have slightly higher I-values. Also in this case, we find that although the Boeing 707 introduced in 1957 can be considered as a dominant design at the firm level, the global diffusion of this design was based on subsequently redesigned models.

*5.2    The results on convergence*

As explained above, the convergence measures are obtained by comparing the entire population of aircraft during a period of five year with a single design introduced thereafter, in order to measure the extent to which a new design has followed the existing practice at the industry level. The results are plotted in *Figure 4*. A cyclic dynamic similar to the previous figure can be observed reflecting the converging to a common scaling heuristics. The curvature is similar to the trends observed in *Figure 3*, but the lower bounds and upper bounds of the convergence values concern different models, as evidenced, for example, by a time-lag of about 15 years between the respective minima.

**FIGURE 4 ABOUT HERE**

The first paradigm based on a four-engined piston propellers configuration, emerged in the late thirties when convergence values started to decrease. It lasted till the mid-sixties. During this period, the lowest values are those of the 1951-Douglas DC6 (*I=0.029* bits), the 1957-Douglas DC7 (*I=0.038* bits). The low convergence values in the 1950s indicate that the scaling trajectory became global in this period.

During the fifties, three aircraft models show high *I*-values indicating that these designs deviated from the paradigmatic heuristics. These models were: *(i)* the first French turbofan powered Caravelle, introduced in 1955, which could fly over 775 km/h, but which failed in the market partly because of its poor range (only 1740 km); *(ii)* The 1951-DHC3 'Otter', which was the first STOL-aircraft introduced in the civil market by DeHavilland Canada; and *(iii)* the 1955-Convair CV 440, which was used for very short distance flights (below 500 km). These three models can be understood as attempts to find market niches outside the trade-off specifications set by the prevailing heuristics. Only the introduction of STOL-aircraft—STOL stands for "Short Take-Off and Landing"—proved a successful strategy as evidenced by the long-standing monopoly of DeHavilland Canada in this particular segment. The monopoly lasted until the introduction of STOL-aircraft by other firms during the seventies and eighties. The success of STOL-aircraft is obviously related to the fact that these aircraft can be used on very short take-off and landing strips, whereas a standard aircraft is useless in such an environment.

The second cycle leads to the paradigm of the two-engine turbofan aircraft which emerged in the early eighties and is currently dominant in civil aircraft. The 1987-Airbus A320 (*I=0.052*), the 1993-Fokker 70 (*I=0.035*), and the 1994-McDonnell Douglas MD90 *(I=0.045)* are among the designs with the lowest values. Although the first successful introduction of turbofan aircraft goes back to the 1957-Boeing 707 which laid the foundation for the Boeing 700-series, it took about 30 years before the industry converged to a common scaling trajectory.

In other words, Boeing has been very successful in partially "monopolising" its design principles because of high rates of learning-by-doing in the large-scale production. A common scaling pattern at the industry level has only emerged after the European-based Airbus penetrated the market. In the early seventies, Boeing's main competitors, McDonnell Douglas and Lockheed, focused on three-engined models in which one engine is integrated into the tail of the aircraft. These models attempted to supply services similar to the Boeing 700-series, but are technically different from Boeing's models.[5] This is evidenced by some high *I*-values during this period, indicating that designs deviated from current practice. The success of these models was only moderate, and very few other firms followed their example. Firms outside the United States focused on specific market segments, such as supersonic speed

---

[5] Except for the three-engine Boeing 727 (1960). Hereafter Boeing abandoned this design concept.

(1969-Concorde) or short-range flights using turboprops (e.g. 1967-Fokker F28, 1972-Embraer 110).

The delay in the establishment of the second technological paradigm can further be explained by the more general observation that the emergence of a paradigm usually takes more time if a new paradigm is competing with an already existing one. Firms that have been successful in the old paradigm will tend to build upon their existing competencies. Since a new technological paradigm includes a new set of design principles, it can be expected to be *competence-destroying*. For this reason, the threat of a new paradigm tends to induce innovation in the older paradigm (Anderson and Tushman 1990). Indeed, some established firms continued the further development of (turbo)propeller technology during the sixties and thereafter, including BAe and Fokker.

## 6. Distinguishing types of innovation

Although the cyclical curvatures in *Figure 3* and *Figure 4* are similar, the time lag in the minimum and maximum values suggests that individual designs have often very different diffusion and convergence values. In other words, those aircraft that had a strong impact on the industry are not necessarily the ones to which previous design have converged, and *vice versa*, a design to which previous design converged, will not necessarily diffuse throughout the industry. Four types of designs can then be distinguished in terms of their diffusion value (high/low) and convergence value (high/low). This classification is summarised in *Table 6*, including some examples drawn from the results on the civil aircraft industry:

**Table 6**
*Classification of product innovation in a diffusion/convergence matrix*

|  | low diffusion I-value (large impact on later design) | high diffusion I-value (small impact on later designs) |
|---|---|---|
| high convergence I-value (deviant from existing designs) | *2. breakthroughs*<br><br>e.g.,<br>Douglas DC4<br>Boeing 377<br>Boeing 767 | *3. failures*<br><br>e.g.,<br>seaplanes<br>biplanes<br>supersonic aircraft |
| low convergence value (following existing designs) | *1. scaling trajectories*<br><br>e.g.,<br>Douglas DC-series<br>McDonnell Douglas MD-series<br>Airbus 300-series | *4. niche –monopolies*<br><br>e.g.,<br>Boeing 747<br>STOL-aircraft |

1. Designs in the south-west quadrant are typically models that were developed during the heyday of a paradigm. They can be considered as piecemeal improvements along a *scaling trajectory*. The models were typically developed between 1945 and 1960 in the case of the first cycle (in particular Douglas' DC-series), and after 1980 during the second paradigm period (Airbus 300-series and McDonnell Douglas MD-series).

2. Designs in the north-east quadrant concern aircraft which were initially quite different from their predecessors, but which were then been copied by many firms after their introduction. They can be considered as *breakthroughs* since these designs were original when compared with their precursors. These designs include the 'success stories' of the Douglas DC4 and the Boeing 767.

3.  The category of designs that is located in the north-east quadrant differed radically from current practice and did not diffuse throughout the industry. These aircraft models can be considered as radical innovations without appeal to other firms (*failures*). Most designs in this quadrant were introduced in the pre-war period, including seaplanes (e.g., Handley Page), biplanes (e.g., DeHavilland) and one-engined aircraft (e.g., Northrop Delta, Lockheed Vega). In the post-war period, these designs concern supersonic aircraft which have been introduced in the late sixties (Concorde, Tupolev 144).

4.  Finally, product designs in the south-east quadrant have not been too different from their precursors, but did not diffuse thereafter. These models concern aircraft that specialized in a niche or obtained a monopoly within a mass-market. Prime examples of designs that succeeded in occupying a niche are the long-range turbofans (Boeing 747, Douglas DC8, Ilyushin IL-86) and DeHavilland Canada STOL-series.

**7.  Comparison with designs for helicopters**

The same methodology was applied to a database containing 180 helicopters (*1940-1996*), which was available from a previous study (Saviotti and Trickett 1993) and which we updated to 1996 for present purposes using Jane's (1998). The only difference between the aircraft and helicopters data is that helicopters are described by the rotor diameter characteristic (in meters) instead of wingspan characteristic in the case of aircraft (also in meters). We used all helicopter types (military and civil), since little differentiation has taken place. The majority of models are used for a variety of purposes (Bilstein 1996).

The results for the diffusion and convergence values of helicopters are plotted in *Figure 5* and *Figure 6*. These results indicate that the technological developments have taken place more smoothly than in aircraft development. After an initial start-up phase during WW II, new helicopter models have continuously been introduced, while there is also ongoing diffusion of previous models. No cyclic dynamic can be discerned at the level of the industry which indicates the absence of a clear pre-paradigmatic and paradigmatic stage in helicopter technology. However, some convergence has been taking place from the sixties onwards, as most helicopters incorporate two turboshaft engines and a single rotor which can be considered as a slowly emerging dominant design. The recent fall in $I$-values may point to an emerging scaling trajectory based on this technology.

FIGURES 5 AND 6 ABOUT HERE

As noted, helicopters are used for a variety of very specific services such as low-altitude fights, ambulance missions, transport of troops, and in off-shore activities. Compared to the aircraft industry, industry sales are low. Furthermore, the market position of helicopters has been further impoverished by competition from STOL-aircraft which have been developed since the fifties. This type of aircraft can take off and land using very short runways allowing it to be used in a variety of natural environments, which has been one of the main advantages of helicopters over other kinds of aircraft. The relatively small number of helicopter sales implies that the impact of learning curves on subsequent technological development is much smaller than it has been the case in the history of aircraft. This may explain why a sudden diffusion of and convergence to a common scaling trajectory, as observed twice in the aircraft industry, has been absent in the history of helicopters. Rather, technological development in helicopter technology takes place gradually and tends to converge only slowly to a common scaling pattern.

## 8. Reconstruction of firm strategies

We have argued that a dominant design at the level of the firm does not necessarily function as a dominant design at the level of the industry. In the cases of both the DC3 and the Boeing 707 we found critical transitions, but the industry analysis showed that these models were not the dominant designs at the level of the industry. The industry was going through a period of reengineering. The diffusion of a dominant design throughout an industry forces firms to choose between following the principles of the emerging dominant design or to continue their own design line (unless a firm itself is at the basis of the dominant design). Given these two dynamics, that is, at the level of the firm and at the level of the industry, one can reconstruct each firm's product sequence either as an elaboration of its own production line of designs or as a shift toward elaborating on the dominant design emerging at the industry level.

These alternative strategies can be tested in the two cases of competition between Boeing and Douglas using the above model of a critical transition. First, the product sequence that Boeing developed during the late thirties after the dominant Douglas DC3 had emerged, can be compared with the assumption of continuity in the firm's own development. Second, the product sequence which McDonnell Douglas developed during the sixties after the dominant Boeing 707 had been introduced, can be assessed in terms of this firm's response to Boeing's breakthrough.

*8.1    The impact of the DC3 on Boeing's firm sequence*

Two major designs were competing on the American market in the 1930s: the Boeing 247 developed in 1930 and the Douglas DC3 first put into service in 1936. The DC3 outcompeted the Boeing 247, as it had both a high speed and range level and allowed for the transport of 21-28 passengers, whereas the Boeing had only 10-12 passenger seats. Within two years after its introduction, the DC3 was used by the majority of airliners (Jane's 1978). Following the stunning success of the DC3, Boeing decided to focus on long-range, four-engine piston propeller aircraft. Boeing developed the 307 Stratoliner in 1938 and the 377 Stratocruiser of 1944 as successors to the Boeing 247. As noted, this design line has been successful in the competition with the DC-line.

The history of Boeing's product sequence can be reconstructed as either having followed up on the its own model, that is, the 247, or as in reaction to the DC3. *Table 7* lists the values of the test for critical transitions for these two possible lines of reconstruction. The first sequence, in which the development of the 307 and the 377 is reconstructed as following up on the Boeing 247D, shows a negative value, which indicates that the Boeing 307 marked a critical transition with respect to Boeing's design practice. The new product line 307 – 377 meant a reorientation with respect to the 247 model. But if the product sequence is reconstructed as a follow-up of the Douglas DC3, the 307 does not exhibit a critical transition. In other words, within a time span of only two years, Boeing's new designs were in line with DC3 which would later become dominant at the industry level in its redesigned DC4-version.

**Table 7**

*Reconstruction of Boeing 307's precursor*

| product sequence | value (in bits) |
| --- | --- |
| Boeing 247D - Boeing 307 - Boeing 377 | -0.0093 |
| Douglas DC3 - Boeing 307 - Boeing 377 | +0.0383 |

*8.2    The impact of the Boeing 707 on Douglas' firm sequence*

After the successful introduction of the turbofan-engine Boeing 707 in 1957, Douglas started in response its own turbofan line in 1967 with the DC8, which was followed by the DC9. Again, one is able to reconstruct the history of this product sequence as heralded by a previous design developed within the firm, that is, the Douglas DC7 of 1956, or as following from the Boeing 707. The results for critical transitions of the two reconstructions are listed in *Table 8*. Like in the previous case, the results show that the switch to a new product sequence marked a critical transition for the firm in question (here, Douglas). However, when the product sequence is reconstructed as a follow-up to the dominant design, the sequence does not show a critical transition.

**Table 8**

*Reconstruction of DC8's precursor*

| product sequence | value (in bits) |
|---|---|
| DC7 - DC8 - DC9 | -0.0176 |
| Boeing 707 - DC8 - DC9 | +0.0091 |

Both reconstructions show that a reorientation at the level of the firm may be triggered by the emergence of a dominant design developed by a competing firm. This reorientation implies an 'unlearning' of design principles incorporated in previous designs, and a re-building of competencies associated with the dominant design. This reorientation is indicated by the critical transitions we found in the trajectory of both firms as a reaction to the success of the competitor.

Note the dynamics of the competition in relation to our results at the industry level. Companies reorient their designs as a reaction to the success of a competitor's design. Reengineered designs then diffuse thereafter throughout the industry. In the case of the first piston propeller paradigm these designs concern the DC4 in 1942 and the Boeing 377 in 1944, respectively (see *Figure 3*). In the second turbofan paradigm, the reengineered two-engine models Boeing 757 and 767 and Airbus 300-310-320 marked the industry-wide diffusion of turbofan technology in the early eighties. Importantly, the global diffusion throughout the industry of a common scaling pattern occurred only after the European competitors had reorganised themselves within the Airbus consortium.

These examples highlight the notion that the relevant level of analysis is expected to change during the product life-cycle (Foray and Garrouste 1991). During the emergence of a dominant design, the analysis of individual trajectories reveals the strategies of firms concerned with building up their competence base. As competing firms are forced to re-build their competencies so as to profit from the learning externalities associated with the dominant design, a single set of design principles can be expected to become established at the industry level. At this point, the relevant dynamics shift from individual design trajectories towards an industrial trajectory.

## 9.     Concluding remarks

We have shown how evolutionary concepts can be addressed in an empirical research design by using methodologies from information calculus. Dynamic distance measures enable us to analyse the data at the level of individual firms, at the industry level, and in relation to each other. At different stages of development we highlighted different dynamics at different levels of aggregation. Our main finding has been that three processes can be distinguished in both product cycles: (i) the emergence of a leading design, (ii) the diffusion process of the scaling patterns throughout the industry, and (iii) the convergence of designs within a technological paradigm towards a common scaling trajectory. Designs that diffused most were not the ones to which the industry had converged before. We found a time lag of about 15 years between processes of diffusion and convergence.

At the firm level, product design sequences may turn out to be critical in reorienting the relation between design variables as each new design changes the probability of which type of design will become dominant within the firm. Once a dominant design is established at the firm level, its design principles are further codified *within* the firm through the development of subsequently scaled designs. We have observed critical transitions associated with the legendary designs of the DC3 and the Boeing 707. Hereafter, the respective firms were able to scale their design into different directions showing the versatility of a dominant design in meeting different user needs. At the industry level, redesigned aircraft models have diffused to the largest extent (DC4/ Boeing 377 and Boeing 767/ Airbus 320). Hereafter, as design principles become increasingly codified *among* firms, the industry can be expected to converge towards common scaling heuristics which make up a technological paradigm (Cowan and Foary 1997). The implication for strategic management holds that the competition changes in character. The diffusion of and convergence to a dominant set of scaling principles implies that price competition on the basis of scaling takes over from product competition (cf. Utterback and Abernathy 1975).

It is important, in our opinion, to distinguish between critical transitions at the micro-level of the firm and the unintended critical transition at the level of the industry because the different stages of the development provide us with other policy and management options. From a consumer welfare perspective, it is crucial to avoid long periods of competing standards between competing firms or nationally supported industries. Supranational policy coordination can be helpful, since competing designs may result in state competition. In this light, the Airbus consortium can be considered as a successful attempt to avoid competing dominant designs at the European level. At the same time, Airbus contributed to the diffusion and further development of turbofan-engine aircraft in the civil market, as pioneered by Boeing.

When an industry is fully committed to a technological paradigm, an R&D-focus on presumptive anomalies can be a sensible strategy. Once radical new design principles begin to diffuse, however, firms and policy-makers are faced with the decision either to improve upon their existing competencies or to re-build their competencies so as to profit from new insights developed by competitors. The former strategy is less costly, but also less ambitious: sticking to an old scaling trajectory will usually force one to accept only a relatively small niche in the longer run. The other strategy is risky since each firm competes by running down its learning curves. However, to follow both strategies may imply an incoherent set of competencies which is costly to maintain.[6] Therefore, a choice must be made at some point in time. Technology policy should assess both the stage of the product life-cycle in a given sector and the ambition level of the industry that one wishes to support.

---

[6] The importance to choose either to exploit existing competencies or to switch to a new set of competencies, can be illustrated by the case of Fokker which recently went bankrupt. The 1955-F.27 incorporated turboprops, while the 1967-F.28 incorporated turbofan engines. This mix was repeated in the eighties when Fokker introduced the F.50 (turboprop) and the F.100 (turbofan), and in the nineties with the development of the F.60 (turboprop) and the F.70 (turbofan).

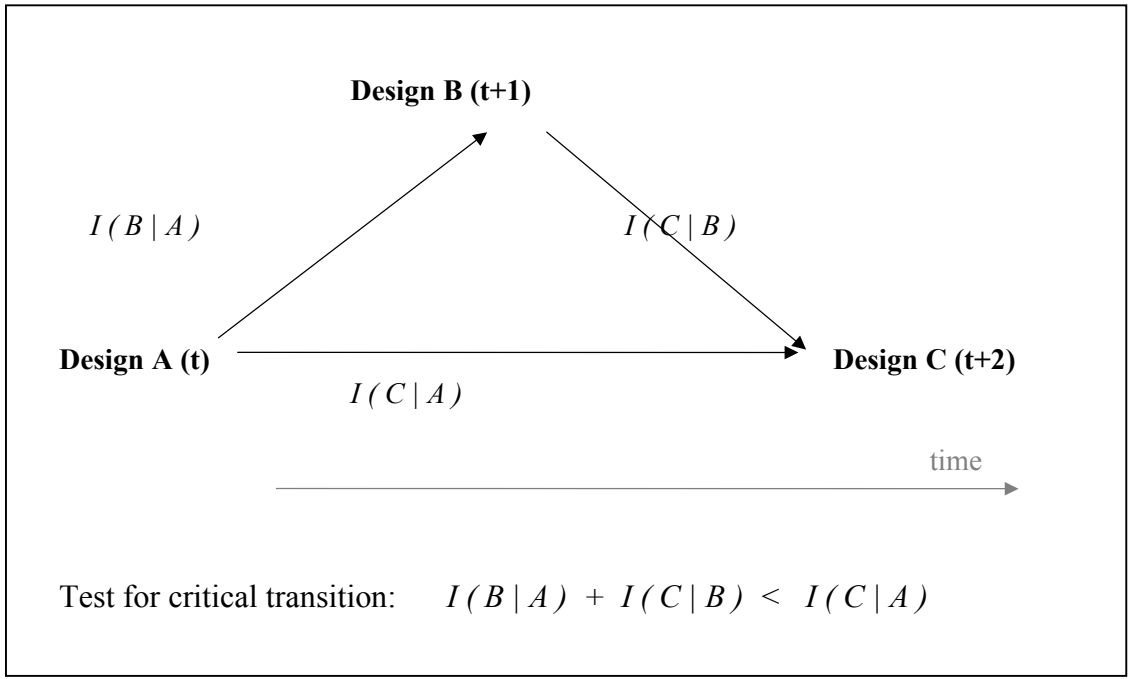

*Figure 1: Schematic representation of a product sequence with informational distances*

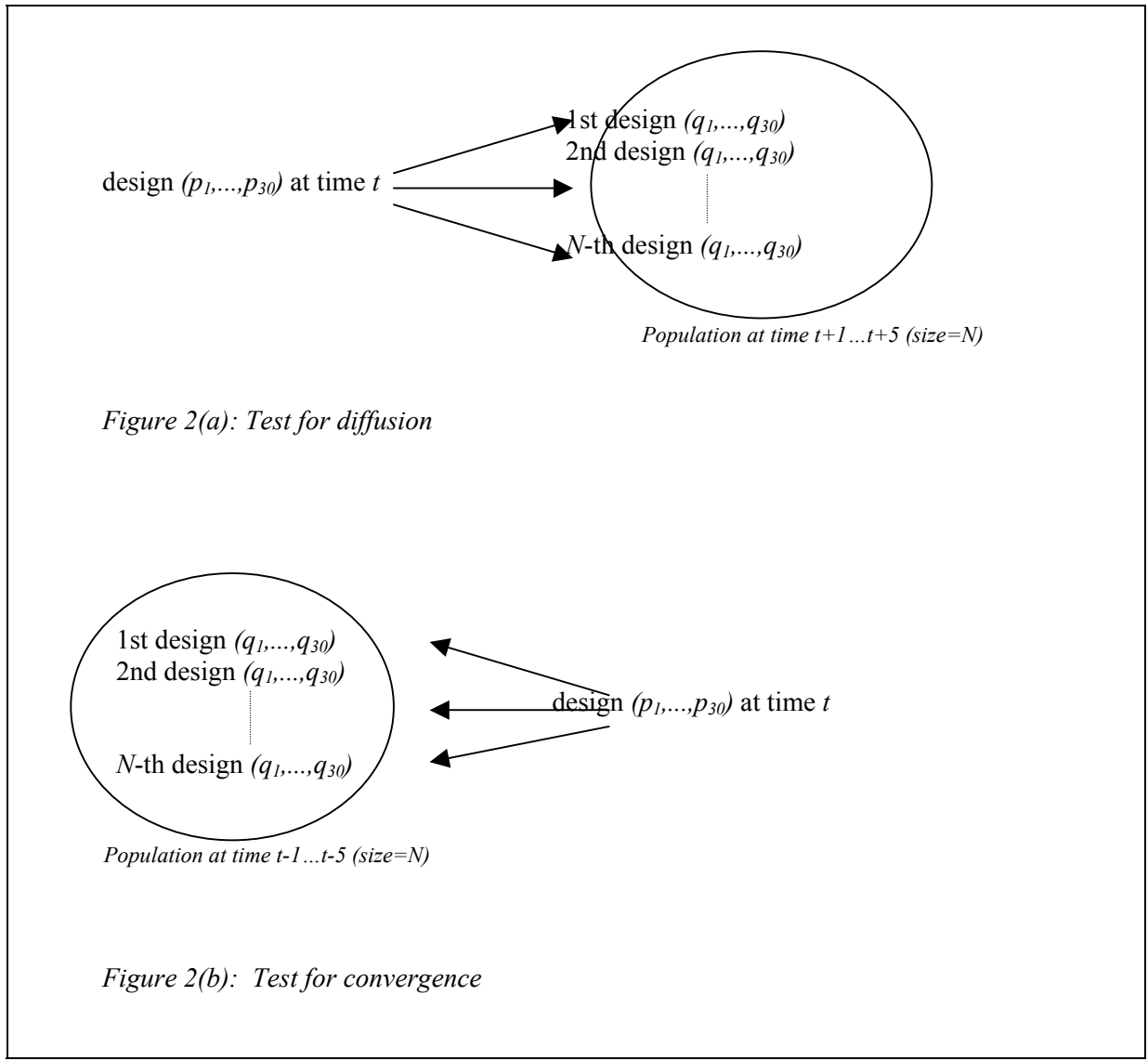

*Figure 2: Schematic representation of diffusion and convergence analysis*

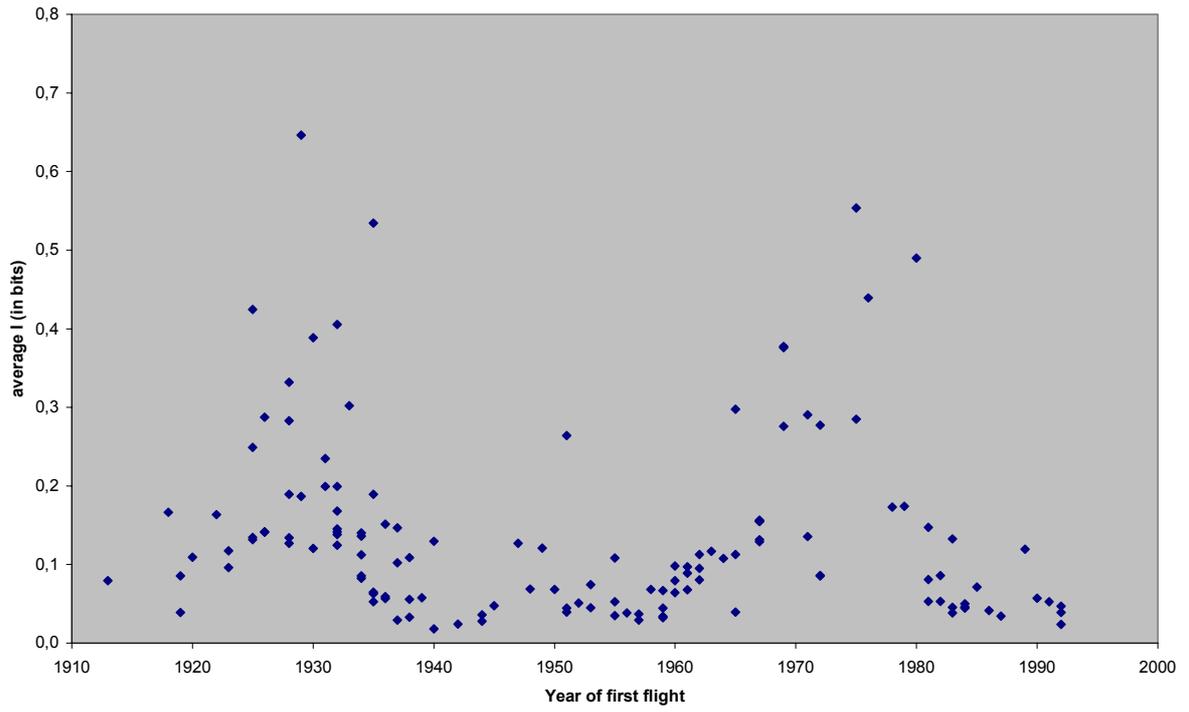

*Figure 3: Diffusion I-values for civil aircraft*

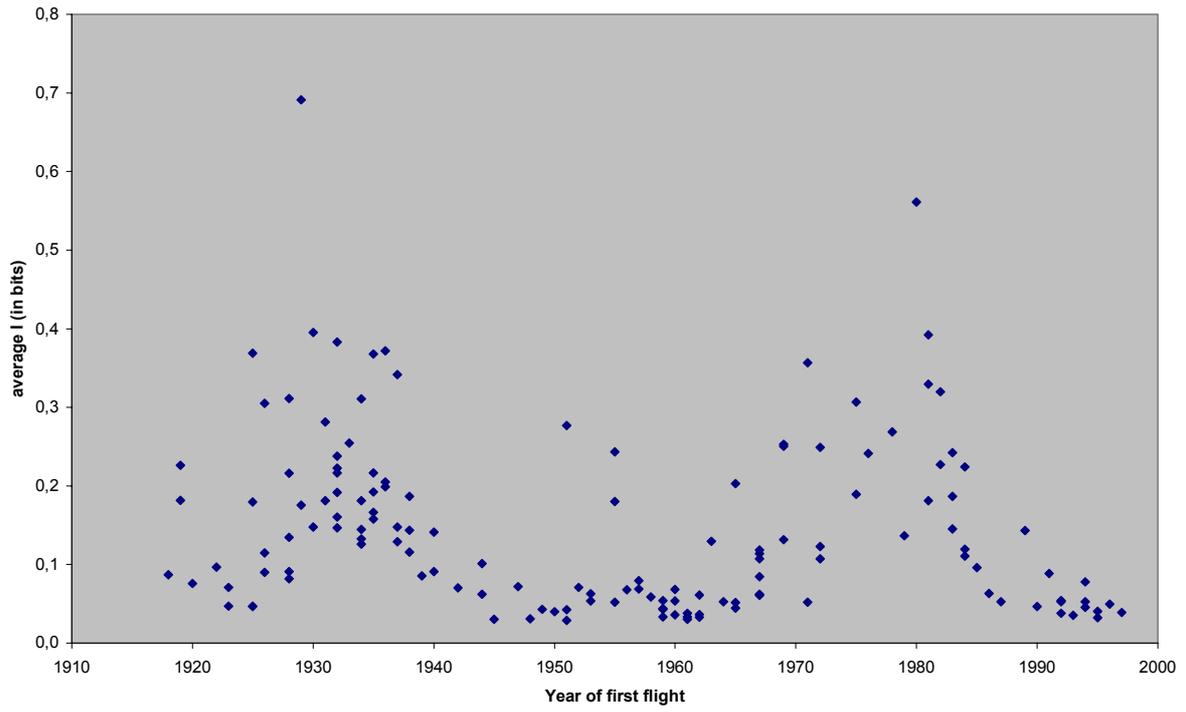

*Figure 4: Convergence I-values for civil aircraft*

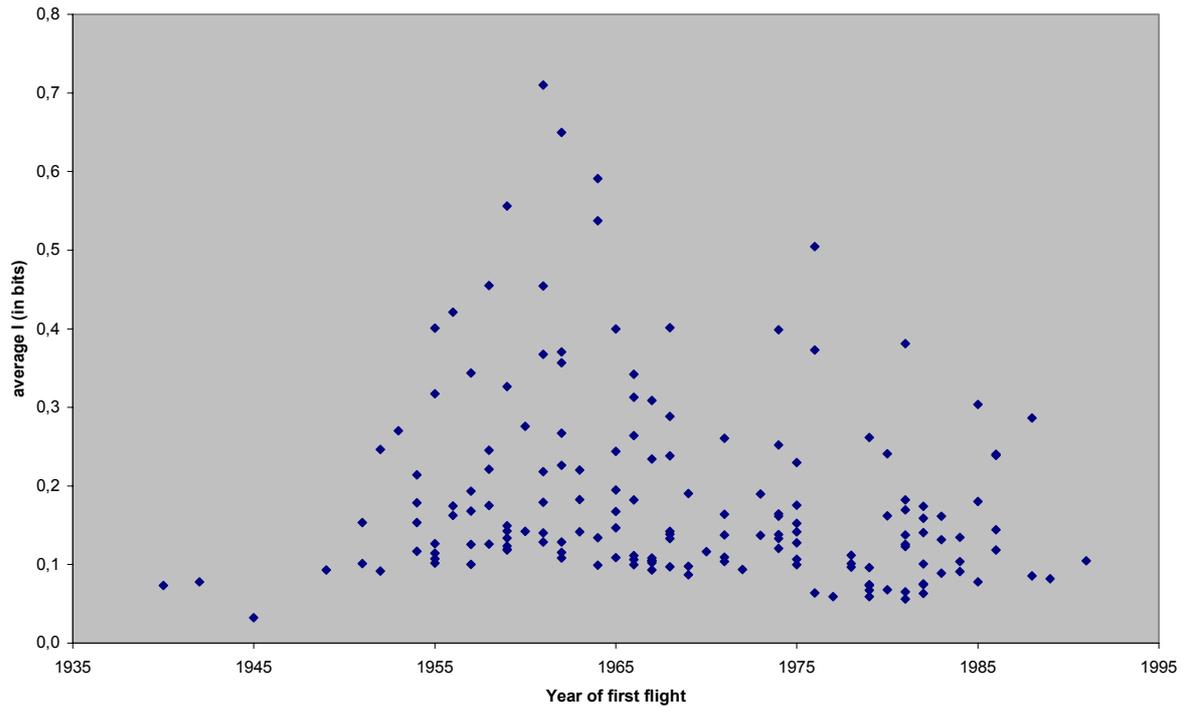

*Figure 5: Diffusion I-values for helicopters*

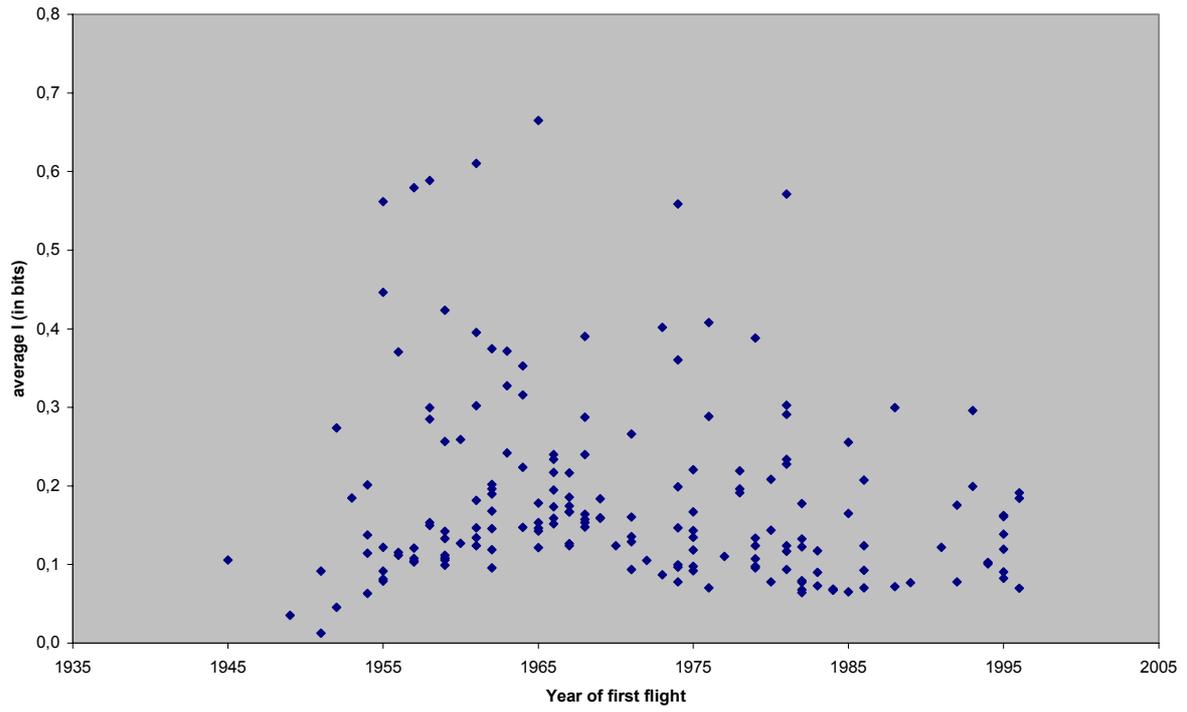

*Figure 6: Convergence I-values for helicopters*